\documentclass[a4paper, oneside, twocolumn, notitlepage, 10pt]{extarticle_ecoc}
\usepackage{ecoc}

\begin{document}
\selectlanguage{english}    


\title{On the Impact of Gas-Line Absorption in Long-Haul C+L-Band Hollow-Core Fibre Transmission}%


\author{
    Zelin~Gan*,
    Eric~Sillekens,
    Jiaqian~Yang,
    Ronit~Sohanpal,
    Mindaugas~Jarmolovi\v{c}ius, \\
    Romulo~Aparecido, 
    Robert~I.~Killey,
    Polina~Bayvel
}

\maketitle                  


\begin{strip}
    \begin{author_descr}

        Optical Networks Group, UCL (University College London), London, UK,
        *\textcolor{blue}{\uline{zelin.gan.17@ucl.ac.uk}}

    \end{author_descr}
\end{strip}

\renewcommand\footnotemark{}
\renewcommand\footnoterule{}


\begin{strip}
    \begin{ecoc_abstract}
        We numerically investigate gas-line absorption in hollow-core fibre C+L-band transmission. Results show limited C-band but severe L-band performance degradation. Ideal suppression enables a 1.5$\times$ higher L-band throughput, reaching 59.1~Tb/s over 1000~km, while reducing repeater number by up to 3.3$\times$. ©2026 The Author(s) 
    \end{ecoc_abstract}
\end{strip}

\section{Introduction}
Hollow-core fibre (HCF) has emerged as a promising alternative to conventional solid-core single-mode fibre (SMF) for future high-capacity optical transmission. HCF offers several intrinsic advantages, including lower Kerr nonlinearity, lower chromatic dispersion, and substantially reduced latency. In addition, development of HCF over the past few years has led to lower attenuation and potentially significantly broader low-loss transmission window than SMF, making it highly attractive for ultra-wideband systems~\cite{petrovich2025broadband}.

However, a major impairment in HCF performance is gas-line absorption (GLA), caused by residual atmospheric gases trapped inside the hollow region, primarily carbon monoxide (CO), carbon dioxide (CO\textsubscript{2}), and water vapour (H\textsubscript{2}O), with CO\textsubscript{2} having greatest impact on performance in the C+L bands~\cite{xiong2025co2}. These absorption features appear as narrow spectral notches, with linewidths typically in the order of 1~GHz, producing frequency-selective fading and signal distortion which leads to inter-symbol interference (ISI), SNR degradation and reduced throughput after long-haul transmission~\cite{he2026accelerating}. Recent work has modelled GLA using HITRAN-based absorption spectra and validated the resulting spectral characteristics against experimental measurements in HCF transmission systems~\cite{gordon2026hitran,chen2025characteristics}. A simplified modelling approach has also been proposed in which the GLA-induced impairment is represented through additive penalty functions that relate the depth and position of each absorption line to the resulting OSNR penalty~\cite{he2026accelerating}. In addition to GLA, inter-modal interference (IMI) from the excitation and propagation of higher-order modes in inherently multi-moded HCF can degrade signal quality and limit transmission distance, particularly in long-haul scenarios~\cite{peng2026low}.

Accordingly, several mitigation approaches have recently been proposed. On the DSP side, spectral pre-equalisation and transmitter pre-emphasis have been shown to effectively compensate part of the CO\textsubscript{2}-induced fading and extend transmission reach~\cite{eric2026gas,li2026dsp}. Waveform-level mitigation has also been explored through multicarrier and entropy-loaded OFDM schemes~\cite{sampaio2025hollow}. Fibre-level solutions such as post-processing with pressurised nitrogen (N\textsubscript{2}) purging have demonstrated complete removal of CO\textsubscript{2} absorption peaks and establishment of stable internal overpressure, although the processing time makes this approach more suitable for short cable sections~\cite{xiong2025co2}. GLA remains a key obstacle to the practical deployment of future C+L-band systems.

In this work, we investigated the impact of GLA on long-haul C+L-band HCF transmission. We numerically modelled HCF transmission including both GLA and IMI, and compare the results against a closed-form SMF benchmark. We have quantified the throughput gain and the reduction in the repeater distance  enabled by GLA suppression, and identify the key limitations for future HCF transmission systems.

\section{Fibre Model}
The simulated attenuation and dispersion profiles of SMF and HCF are shown in Fig.~\ref{fig:att_disp}. For SMF, the wavelength-dependent attenuation, dispersion, and Raman gain spectrum reported in~\cite{Min_JLT} are used. For HCF, the total attenuation is modelled as the combination of a baseline fibre attenuation, approximately 0.07~dB/km across the C+L-band, and GLA~\cite{petrovich2025broadband}. The GLA in HCF is modelled using spectroscopic parameters from the HITRAN database. For each molecular transition, the absorption line shape is represented by a Voigt profile. The absorption coefficient is given by
\begin{equation}
    \alpha(\nu)=\sum\nolimits_i S_i\,V(\nu-\nu^*_i;\sigma_i,\gamma_i),
\end{equation}
where $S_i$ is the temperature-dependent line strength, $\nu^*_i$ is the pressure-shifted line-center frequency, and $V$ denotes the Voigt profile with Gaussian width $\sigma_i$ determined by Doppler broadening and Lorentzian width $\gamma_i$ determined by pressure broadening. The transmission absorption through distance $L$ was then calculated as $|H_{\mathrm{GLA}}(\nu)|=e^{-\alpha(\nu)L/2}$. We further assumed a minimum-phase response, such that the phase is related to the amplitude through the Hilbert-transform relation,
\begin{equation}
\phi(\nu)=-\mathcal{H}\left\{\ln |H_{\mathrm{GLA}}(\nu)|\right\},
\end{equation}
and the complex transfer function becomes
\begin{equation}
H_{\mathrm{GLA}}(\nu)=|H_{\mathrm{GLA}}(\nu)|e^{j\phi(\nu)}.
\end{equation}
In Fig.~\ref{fig:att_disp}, the benchmark GLA case is modelled with a core pressure of 0.2~bar~\cite{xiong2025co2} and 1000~ppm CO\textsubscript{2} concentration at a temperature of 296~K. The GLA suppression is modelled as the gas purging approach with varying the CO\textsubscript{2} concentration or equivalently the partial pressure of CO\textsubscript{2}.

\pgfplotstableread{Data/fibre_hcf_np_gla.txt}\glaTable
\pgfplotstablegetrowsof{\glaTable}
\pgfmathtruncatemacro{\LastRow}{\pgfplotsretval-1}

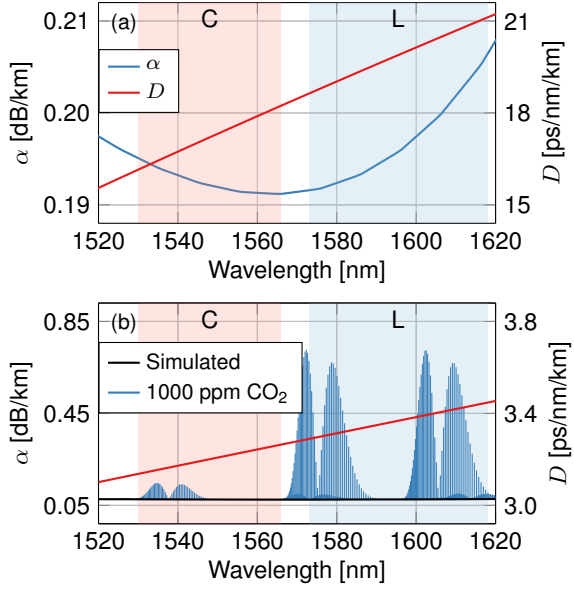
\begin{figure}[!t]
\centering
    \begin{tikzpicture}\small
        \begin{groupplot}[
            group style={group size=1 by 4,vertical sep=30pt},
            width=\linewidth-22.71675pt,
            height=4.5cm,
            xmin=1520, xmax=1620,
            xtick={1520,1540,1560,1580,1600,1620},
            xticklabels=\empty,
            xlabel near ticks,
            ylabel near ticks,
            ylabel shift = -2 pt,
            xlabel shift = -2 pt,
            grid=both,
            unbounded coords=jump,
            set layers
        ]
        \nextgroupplot[
            axis y line*=left,
            xlabel={Wavelength [nm]},
            ylabel={$\alpha$~[dB/km]},
            xticklabels={1520,1540,1560,1580,1600,1620},
            ymin=0.188,
            ymax=0.212,
            ytick={0.19,0.20,0.21},
            yticklabels={0.19,0.20,0.21},
            legend style={/tikz/font=\footnotesize,anchor=south west,at={(axis cs: 1520,0.2)}},
            legend cell align={left},
            legend columns=1,
            legend image post style={xscale=0.7},
        ]
        \addplot[Set1-B,thick,solid,mark=none] table[x=wavelength,y=attenuation] {Data/fibre.txt};
        \fill[Set3-D,opacity=0.2] (axis cs:1530,0.188) rectangle (axis cs:1566,0.212);
        \fill[Set3-E,opacity=0.2] (axis cs:1573,0.188) rectangle (axis cs:1618,0.212);
        \node[anchor=mid] at (axis cs:1548,0.21){C};
        \node[anchor=mid] at (axis cs:1595.5,0.21){L};

        \addlegendimage{Set1-A,thick,solid,mark=none};
        \addlegendentry{$\alpha$};
        \addlegendimage{Set1-B,thick,solid,mark=none};
        \addlegendentry{$D$};
        
        \nextgroupplot[at=(group c1r1.north),
            axis y line*=right,
            axis x line=none,
            ylabel={$D$~[ps/nm/km]},
            ymin=14.4,
            ymax=21.6,
            ytick={15,18,21},
            yticklabels={15,18,21},
        ]
        \addplot[Set1-A,thick,solid,mark=none] table[x=wavelength,y=dispersion] {Data/fibre.txt};

        \node[anchor=north west,font=\footnotesize] at (rel axis cs:0.01,0.99) {(a)};
        
        \nextgroupplot[
            axis y line*=left,
            ylabel={$\alpha$~[dB/km]},
            ymin=-0.03,
            ymax=0.93,
            ytick={0.05,0.45,0.85},
            yticklabels={0.05,0.45,0.85},
            legend style={/tikz/font=\footnotesize,anchor=north west,at={(axis cs: 1520,0.15)}},
            legend cell align={left},
            legend columns=2,
        ]
        \addplot[black,thick,solid,mark=none,forget plot,on layer=axis foreground] table[x=wav,y=att] {Data/fibre_hcf_np.txt};
        \fill[Set3-D,opacity=0.2] (axis cs:1530,-0.03) rectangle (axis cs:1566,0.93);
        \fill[Set3-E,opacity=0.2] (axis cs:1573,-0.03) rectangle (axis cs:1618,0.93);
        \node[anchor=mid] at (axis cs:1548,0.85){C};
        \node[anchor=mid] at (axis cs:1595.5,0.85){L};
        \foreach \r in {0,...,\LastRow} {
            \pgfplotstablegetelem{\r}{gl}\of{\glaTable}
            \let\xval\pgfplotsretval
            \pgfplotstablegetelem{\r}{gla_b}\of{\glaTable}
            \let\ybot\pgfplotsretval
            \pgfplotstablegetelem{\r}{gla_t}\of{\glaTable}
            \let\ytop\pgfplotsretval
        
            \addplot[
                Set1-B,
                mark=none,
                forget plot
            ] coordinates {
                (\xval,\ybot)
                (\xval,\ytop)
            };
        }
        \nextgroupplot[at=(group c1r3.north),
            axis y line*=right,
            ylabel={$D$~[ps/nm/km]},
            ymin=2.92,
            ymax=3.88,
            ytick={3.0,3.4,3.8},
            yticklabels={3.0,3.4,3.8},
            xticklabels={1520,1540,1560,1580,1600,1620},
            xticklabel style={/pgf/number format/1000 sep=},
            xlabel={Wavelength [nm]},
            legend style={/tikz/font=\footnotesize,anchor=south west,at={(axis cs: 1520,3.4)},cells={align=center}},
            legend cell align={left},
            legend columns=1,
            legend image post style={xscale=0.7},
        ]
        \addplot[Set1-A,thick,solid,mark=none,forget plot] table[x=wav,y=disp] {Data/fibre_hcf_np.txt};
        
        \addlegendimage{black,thick,solid,mark=none};
        \addlegendentry{Simulated};
        \addlegendimage{Set1-B,thick,solid,mark=none};
        \addlegendentry{1000~ppm CO\textsubscript{2}};

        \node[anchor=north west,font=\footnotesize] at (rel axis cs:0.01,0.99) {(b)};
        
        \end{groupplot}
        
    \end{tikzpicture}
    \caption{Simulated attenuation $\alpha(\lambda)$ and dispersion $D(\lambda)$ of (a) SMF and (b) HCF.}
    \label{fig:att_disp}
\end{figure}
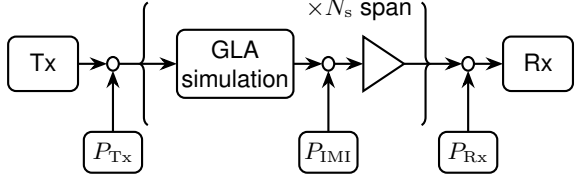
\begin{figure}
\centering
\begin{tikzpicture}[xscale=0.77,yscale=0.9]
    \newcommand{\loadnoise}[3]{
    \begin{scope}[shift={(#1,#2)}]
        \draw[thick,rounded corners=3] (0,0) rectangle (1,0.6);
        \node[anchor=mid,font=\small] at (0.5,0.3) {#3};
        \draw[-Stealth,thick] (0.5,0.6) -- (0.5,1.5);
    \end{scope}%
    }
    \draw[thick,rounded corners=3] (0.4,0) rectangle (1.6,0.8);
    \node[anchor=mid,font=\small] at (1.0,0.4) {Tx};
    \draw[-Stealth,thick] (1.6,0.4) -- (2.1,0.4);
    \draw[thick] (2.2,0.4) circle [radius=0.1];
    \loadnoise{2.2-0.5}{0.3-1.5}{$P_{\mathrm{Tx}}$};
    \draw[-Stealth,thick] (2.3,0.4) -- (3.3,0.4);
    \draw[decorate,decoration={brace,amplitude=3},thick] (2.8,0.4-0.9) -- (2.8,0.4+0.9);
    \draw[decorate,decoration={brace,amplitude=3,mirror},thick] (7.5,0.4-0.9) -- (7.5,0.4+0.9);
    \node[anchor=north east,font=\small] at (7.5,0.4+0.9+0.2) {$\times N_{\mathrm{s}}$ span};
    \draw[thick,rounded corners=3] (3.3,-0.1) rectangle (5.3,0.9);
    \node[anchor=center,font=\small,align=center] at (4.3,0.4) {GLA \\ simulation};
    \draw[-Stealth,thick] (5.3,0.4) -- (5.8,0.4);
    \draw[thick] (5.9,0.4) circle [radius=0.1];
    \loadnoise{5.9-0.5}{0.3-1.5}{$P_{\mathrm{IMI}}$};
    \draw[-Stealth,thick] (6.0,0.4) -- (6.5,0.4);
    \draw[thick,fill=white] (6.5,0.0) -- (6.5,0.8) -- (6.5+0.7,0.4) -- cycle;
    \draw[-Stealth,thick] (7.2,0.4) -- (8.2,0.4);
    \draw[thick] (8.3,0.4) circle [radius=0.1];
    \loadnoise{8.3-0.5}{0.3-1.5}{$P_{\mathrm{Rx}}$};
    \draw[-Stealth,thick] (8.4,0.4) -- (8.9,0.4);
    \draw[thick,rounded corners=3] (8.9,0) rectangle (10.1,0.8);
    \node[anchor=mid,font=\small] at (9.5,0.4) {Rx};

    \path[use as bounding box]
        ([xshift=-2pt,yshift=-2pt]current bounding box.south west)
        rectangle
        ([xshift=2pt,yshift=0pt]current bounding box.north east);
\end{tikzpicture}
\caption{Schematics of the HCF simulation.}
\label{fig:setup}

\end{figure}
\begin{figure}
\centering
\begin{tikzpicture}\small
    \begin{groupplot}[
        group style={group size=1 by 3,vertical sep=30pt,xlabels at=edge bottom,ylabels at=edge left},
        grid=both,
        width=\linewidth-14.91649pt,
        height=4.5cm,
        ylabel near ticks,
        xlabel near ticks,
        xlabel shift = -2 pt,
        ylabel shift = -2 pt,
        xtick=\empty,
        xticklabels=\empty,
    ]
    \nextgroupplot[
        legend columns=1,
        legend style={fill opacity=1,draw opacity=1,text opacity=1,at={(0.0,0.0)},anchor=south west,draw=black,/tikz/font=\footnotesize},
        legend cell align={left},
        xlabel={Wavelength~[nm]},
        ylabel={SNR~[dB]},
        ymin=11.2,
        ymax=24.8,
        ytick={12,16,20,24},
        yticklabels={12,16,20,24},
        xmin=1520,
        xmax=1620,
        xtick={1520,1540,1560,1580,1600,1620},
        xticklabels={1520,1540,1560,1580,1600,1620},
    ]
        \fill[Set3-D,opacity=0.2] (axis cs:1530,11.2) rectangle (axis cs:1566,24.8);
        \fill[Set3-E,opacity=0.2] (axis cs:1573,11.2) rectangle (axis cs:1618,24.8);
        \node[anchor=mid] at (axis cs:1548,24){C};
        \node[anchor=mid] at (axis cs:1595.5,24){L};
        \addplot[black,thick,solid,only marks,mark=*,mark size=0.5pt] table[x=wavelength_nm,y=smf_snr_db] {Data/snr_all.txt};
        \addlegendentry{SMF};
        \addplot[Set1-A,thick,solid,only marks,mark=*,mark size=0.5pt] table[x=wavelength_nm,y=hcf_snr_gla_100] {Data/snr_all.txt};
        \addlegendentry{1000~ppm CO\textsubscript{2}};
        \addplot[Set1-B,thick,solid,only marks,mark=*,mark size=0.5pt] table[x=wavelength_nm,y=hcf_snr_gla_040] {Data/snr_all.txt};
        \addlegendentry{400~ppm CO\textsubscript{2}};

        \node[anchor=north east,font=\footnotesize] at (rel axis cs:0.99,0.99) {(a)};

    \nextgroupplot[
        axis y line*=left,
        legend columns=1,
        legend style={fill opacity=1,draw opacity=1,text opacity=1,at={(0.0,0.56)},anchor=south west,draw=black,/tikz/font=\footnotesize},
        legend cell align={left},
        legend image post style={xscale=0.7},
        xlabel={Frequency~[GHz]},
        ylabel={$|H|^2$~[dB]},
        xmin=-16, xmax=16,
        ymin=-64, ymax=4,
        ytick={-60,-40,-20,0},
        yticklabels={-60,-40,-20,0},
        xtick={-16,-8,0,8,16},
        xticklabels={-16,-8,0,8,16},
    ]
        \addplot[Set1-A,thick,solid,mark=none] table[x=freq,y=mag_dB_100,forget plot] {Data/hcf_tf.txt};
        \addplot[Set1-B,thick,solid,mark=none] table[x=freq,y=mag_dB_040,forget plot] {Data/hcf_tf.txt};

        \addlegendimage{black,thick,solid,mark=none};
        \addlegendentry{Amplitude};
        \addlegendimage{black,thick,dashed,mark=none};
        \addlegendentry{Phase};

        \draw [solid,thick] (axis cs:2.8-1,-53.8)--(axis cs:2.8,-53.8);
        \draw [solid,thick] (axis cs:2.8-1,-21.5)--(axis cs:2.8,-21.5);
        \draw [stealth-stealth,thick] (axis cs:2.8-0.5,-53.8)--(axis cs:2.8-0.5,-21.5) node[midway,fill opacity=0,text opacity=1,inner sep=1pt,font=\footnotesize,anchor=east] {32.3~dB};

    \nextgroupplot[at=(group c1r2.north),
        axis y line*=right,
        legend columns=1,
        legend style={fill opacity=1,draw opacity=1,text opacity=1,at={(0.0,0.0)},anchor=south west,draw=black,/tikz/font=\footnotesize},
        legend cell align={left},
        legend image post style={xscale=0.7},
        ylabel={Phase~[rad]},
        xmin=-16, xmax=16,
        ymin=-6.8, ymax=6.8,
        ytick={-6,-2,2,6},
        yticklabels={-6,-2,2,6},
        ymajorgrids=false,
    ]
        \addplot[Set1-A,thick,dashed,mark=none] table[x=freq,y=phase_rad_100,forget plot] {Data/hcf_tf.txt};
        \addplot[Set1-B,thick,dashed,mark=none] table[x=freq,y=phase_rad_040,forget plot] {Data/hcf_tf.txt};

        \addlegendimage{Set1-A,thick,solid,mark=none};
        \addlegendentry{1000~ppm CO\textsubscript{2}};
        \addlegendimage{Set1-B,thick,solid,mark=none};
        \addlegendentry{400~ppm CO\textsubscript{2}};

        \node[anchor=south east,font=\footnotesize] at (rel axis cs:0.99,0.01) {(b)};
        
    \end{groupplot}
    
\end{tikzpicture}

\caption{(a) Received SNR after first span of a 10$\times$100~km transmission and (b) GLA transfer function.}
\label{fig:snr_tf}

\end{figure}
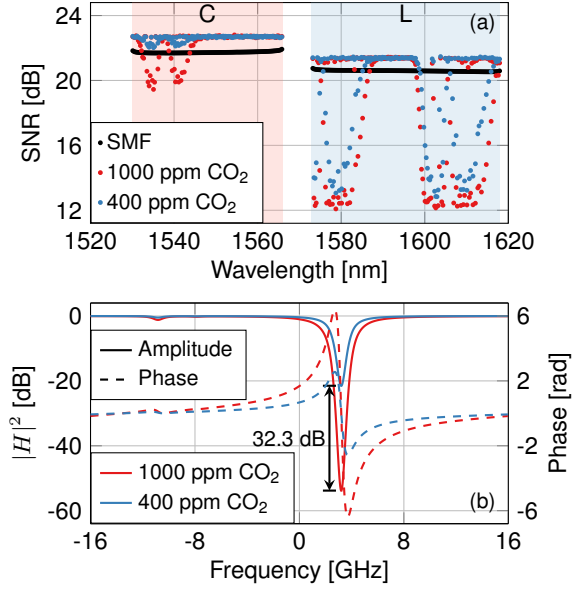

\section{Simulation Setup}
Simulations were carried out using 294 WDM channels with 32~GBaud symbol rate and 33~GHz channel spacing in the C+L-band, with a total of 135 channels in the C-band and 159 channels in the L-band. A flat transceiver SNR of 23.5~dB was assumed in the C-band and 22~dB in the L-band. At the end of each span, ideal amplification was assumed for both the simulated SMF and HCF attenuation, with flat amplifier noise figures of 5~dB in the C-band and 5.5~dB in the L-band.

In the SSMF link, the total received SNR for the $i$-th channel of interest (COI) after one span can be estimated as 
\begin{align}\label{eq:snr}
    \mathrm{SNR}_i &\approx \frac{P_i}{\kappa_{\mathrm{TRx},i}P_i+P_{\mathrm{ASE},i}+P_{\mathrm{NLI},i}},
\end{align}
where $P_i$ is the launch power of the COI, and $\kappa_{\mathrm{TRx},i}=1/\mathrm{SNR}_{\mathrm{TRx},i}$ denotes the transceiver-limited SNR. $P_{\mathrm{ASE},{i}}$ is the ASE noise power at the COI, and $P_{\mathrm{NLI},{i}} = \eta_{\mathrm{NLI}}(f_i)P_i^3$ is the nonlinear interference (NLI) noise power, where $\eta_{\mathrm{NLI}}(f_i)$ is NLI coefficient calculated using the closed-form model in~\cite{lowlossBuglia} for Gaussian-modulated symbols.

In the HCF link, the total received SNR was calculated numerically using the simulation setup in Fig.~\ref{fig:setup}. Gaussian-modulated symbols are generated at the transmitter. The transceiver noise is equally split into transmitter and receiver noise, denoted by $P_\mathrm{Tx}$ and $P_\mathrm{Rx}$, which are added to the signal before and after the transmission. In each span, the GLA transfer function is applied to the signal followed by an additive white IMI noise, $P_\mathrm{IMI}=\kappa_\mathrm{IMI}P_\mathrm{tot}L_\mathrm{s}$ where $\kappa_\mathrm{IMI}$ is the IMI strength, assumed to be -52~dB/km, $P_\mathrm{tot}$ is the total launch power, and $L_\mathrm{s}$ is the span length. The amplifier then compensates for the span loss and adds the corresponding ASE noise. Polarisation-mode dispersion (PMD) and fibre nonlinearity of HCF are neglected in the simulation.

The throughput was estimated using the Shannon information rate, $T_i \approx 2 R_\mathrm{sym}\log_{2}\left(1+\mathrm{SNR}_i\right)$, where $R_\mathrm{sym}$ denotes the symbol rate.

\section{Results}

\begin{figure*}
\centering
\begin{tikzpicture}\small
    \begin{groupplot}[
        group style={group size=2 by 1,horizontal sep=40pt,xlabels at=edge bottom},
        grid=both,
        width=.52\linewidth-7.74338/2pt,
        height=6cm,
        ylabel near ticks,
        xlabel near ticks,
        ylabel shift = -2 pt,
        xlabel shift = -2 pt,
        xlabel={Span length~[km]},
        ylabel={Throughput~[Tb/s]},
        ymin=35,
        ymax=60,
        ytick={35,40,45,50,55,60},
        yticklabels={35,40,45,50,55,60},
        xmin=50,
        xmax=350,
        xtick={50,100,150,200,250,300,350},
        xticklabels={50,100,150,200,250,300,350},
    ]
    \newcommand\marksize{1pt}
    \nextgroupplot[
        legend columns=8,
        legend style={fill opacity=1,draw opacity=1,text opacity=1,at={(0.0,1.04)},anchor=south west,draw=black,/tikz/font=\footnotesize},
        legend cell align={left},
        legend image post style={xscale=0.7},
    ]
        \addplot[Set1-A,thick,solid,mark=o,mark size=\marksize,forget plot] table[x=span_length_km,y=P027_gla100] {Data/throughput_C_all_gla_all_power.txt};
        \addplot[Set1-A,thick,dashed,mark=square,mark size=\marksize,forget plot] table[x=span_length_km,y=P032_gla100] {Data/throughput_C_all_gla_all_power.txt};
        \addplot[Set1-A,thick,dotted,mark=triangle,mark size=\marksize,forget plot] table[x=span_length_km,y=P036_gla100] {Data/throughput_C_all_gla_all_power.txt};
        \addplot[Set1-A,thick,dashdotted,mark=diamond,mark size=\marksize,forget plot] table[x=span_length_km,y=P040_gla100] {Data/throughput_C_all_gla_all_power.txt};

        \addplot[Set1-B,thick,solid,mark=o,mark size=\marksize,forget plot] table[x=span_length_km,y=P027_gla040] {Data/throughput_C_all_gla_all_power.txt};
        \addplot[Set1-B,thick,dashed,mark=square,mark size=\marksize,forget plot] table[x=span_length_km,y=P032_gla040] {Data/throughput_C_all_gla_all_power.txt};
        \addplot[Set1-B,thick,dotted,mark=triangle,mark size=\marksize,forget plot] table[x=span_length_km,y=P036_gla040] {Data/throughput_C_all_gla_all_power.txt};
        \addplot[Set1-B,thick,dashdotted,mark=diamond,mark size=\marksize,forget plot] table[x=span_length_km,y=P040_gla040] {Data/throughput_C_all_gla_all_power.txt};

        \addplot[Set1-C,thick,solid,mark=*,mark size=\marksize,forget plot] table[x=span_length_km,y=P027_gla000] {Data/throughput_C_all_gla_all_power.txt};
        \addplot[Set1-C,thick,dashed,mark=square*,mark size=\marksize,forget plot] table[x=span_length_km,y=P032_gla000] {Data/throughput_C_all_gla_all_power.txt};
        \addplot[Set1-C,thick,dotted,mark=triangle*,mark size=\marksize,forget plot] table[x=span_length_km,y=P036_gla000] {Data/throughput_C_all_gla_all_power.txt};
        \addplot[Set1-C,thick,dashdotted,mark=diamond*,mark size=\marksize,forget plot] table[x=span_length_km,y=P040_gla000] {Data/throughput_C_all_gla_all_power.txt};

        \addplot[Set1-D,thick,solid,mark=pentagon*,mark size=\marksize,forget plot] table[x=span,y=c] {Data/throughput_smf.txt};

        \addlegendimage{Set1-D,thick,solid,mark=pentagon*,mark size=\marksize};
        \addlegendentry{SMF};
        \addlegendimage{Set1-A,thick,solid,mark=none};
        \addlegendentry{1000~ppm CO\textsubscript{2}};
        \addlegendimage{Set1-B,thick,solid,mark=none};
        \addlegendentry{400~ppm CO\textsubscript{2}};
        \addlegendimage{Set1-C,thick,solid,mark=none};
        \addlegendentry{0~ppm CO\textsubscript{2}};
        \addlegendimage{black,thick,solid,mark=*,mark size=\marksize};
        \addlegendentry{27~dBm};
        \addlegendimage{black,thick,dashed,mark=square*,mark size=\marksize};
        \addlegendentry{32~dBm};
        \addlegendimage{black,thick,dotted,mark=triangle*,mark size=\marksize};
        \addlegendentry{36~dBm};
        \addlegendimage{black,thick,dashdotted,mark=diamond*,mark size=\marksize};
        \addlegendentry{40~dBm};

        \draw [stealth-stealth,thick] (axis cs:100,42.2)--(axis cs:270,42.2) node[midway,fill opacity=0,text opacity=1,inner ysep=2pt,font=\footnotesize,anchor=north] {2.7$\times$};

        \draw [stealth-stealth,thick] (axis cs:90,45.6)--(axis cs:90,50.2) node[above,fill opacity=0,text opacity=1,inner ysep=12pt,font=\footnotesize,anchor=south] {4.6~Tb/s};

        \node[anchor=south west,font=\footnotesize] at (rel axis cs:0.01,0.01) {(a)};

    \nextgroupplot[
        legend columns=1,
        legend style={fill opacity=1,draw opacity=1,text opacity=1,at={(0.0,0.0)},anchor=south west,draw=black,/tikz/font=\footnotesize},
        legend cell align={left},
    ]
        \addplot[Set1-A,thick,solid,mark=o,mark size=\marksize,forget plot] table[x=span_length_km,y=P027_gla100] {Data/throughput_L_all_gla_all_power.txt};
        \addplot[Set1-A,thick,dashed,mark=square,mark size=\marksize,forget plot] table[x=span_length_km,y=P032_gla100] {Data/throughput_L_all_gla_all_power.txt};
        \addplot[Set1-A,thick,dotted,mark=triangle,mark size=\marksize,forget plot] table[x=span_length_km,y=P036_gla100] {Data/throughput_L_all_gla_all_power.txt};
        \addplot[Set1-A,thick,dashdotted,mark=diamond,mark size=\marksize,forget plot] table[x=span_length_km,y=P040_gla100] {Data/throughput_L_all_gla_all_power.txt};

        \addplot[Set1-B,thick,solid,mark=o,mark size=\marksize,forget plot] table[x=span_length_km,y=P027_gla040] {Data/throughput_L_all_gla_all_power.txt};
        \addplot[Set1-B,thick,dashed,mark=square,mark size=\marksize,forget plot] table[x=span_length_km,y=P032_gla040] {Data/throughput_L_all_gla_all_power.txt};
        \addplot[Set1-B,thick,dotted,mark=triangle,mark size=\marksize,forget plot] table[x=span_length_km,y=P036_gla040] {Data/throughput_L_all_gla_all_power.txt};
        \addplot[Set1-B,thick,dashdotted,mark=diamond,mark size=\marksize,forget plot] table[x=span_length_km,y=P040_gla040] {Data/throughput_L_all_gla_all_power.txt};

        \addplot[Set1-C,thick,solid,mark=*,mark size=\marksize,forget plot] table[x=span_length_km,y=P027_gla000] {Data/throughput_L_all_gla_all_power.txt};
        \addplot[Set1-C,thick,dashed,mark=square*,mark size=\marksize,forget plot] table[x=span_length_km,y=P032_gla000] {Data/throughput_L_all_gla_all_power.txt};
        \addplot[Set1-C,thick,dotted,mark=triangle*,mark size=\marksize,forget plot] table[x=span_length_km,y=P036_gla000] {Data/throughput_L_all_gla_all_power.txt};
        \addplot[Set1-C,thick,dashdotted,mark=diamond*,mark size=\marksize,forget plot] table[x=span_length_km,y=P040_gla000] {Data/throughput_L_all_gla_all_power.txt};

        \addplot[Set1-D,thick,solid,mark=pentagon*,mark size=\marksize,forget plot] table[x=span,y=l] {Data/throughput_smf.txt};

        \draw [stealth-stealth,thick] (axis cs:60,39.5)--(axis cs:60,57.4) node[midway,fill opacity=0,text opacity=1,inner xsep=1pt,font=\footnotesize,anchor=west] {1.5$\times$};


        \draw [stealth-stealth,thick] (axis cs:100,48)--(axis cs:333,48) node[midway,fill opacity=0,text opacity=1,inner ysep=2pt,font=\footnotesize,anchor=south] {3.3$\times$};

        \draw [-stealth,thick,rotate=0] (axis cs:200,38.5) arc [start angle=-150,end angle=150,x radius=0.1cm,y radius=0.25cm] node[near start,fill opacity=0,text opacity=1,inner ysep=2pt,font=\footnotesize,anchor=north] {Higher power};

        \draw [stealth-stealth,thick] (axis cs:90,40.2)--(axis cs:90,45.5) node[near start,fill opacity=0,text opacity=1,inner ysep=12pt,font=\footnotesize,anchor=north] {5.3~Tb/s};
        
        \node[anchor=south west,font=\footnotesize] at (rel axis cs:0.01,0.01) {(b)};

    \end{groupplot}
    
\end{tikzpicture}

\caption{Achieved throughput from (a) C- and (b) L-band 1000~km SMF and HCF transmission with different residual gas conditions, total launch power and span length.}
\label{fig:throughput_gla_span}

\end{figure*}
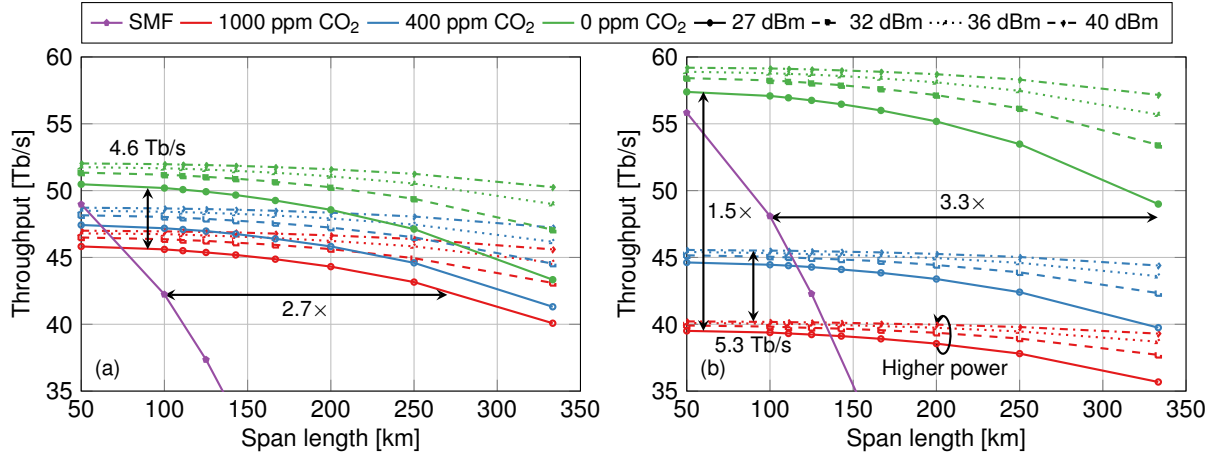

The SNR profiles of SMF and HCF for the first span in 10$\times$100~km transmission are shown in Fig.~\ref{fig:snr_tf}(a). For SMF, the optimal launch powers 20/21~dBm for C/L-band are used, while for HCF 27~dBm is used for both C- and L-band. For HCF, two residual gas conditions are considered: 0 and 400~ppm CO\textsubscript{2}. Channels unaffected by GLA are primarily limited by transceiver noise. Reducing the residual CO\textsubscript{2} concentration to 400~ppm improves the SNR of GLA-affected channels by a mean of 2.3~dB and a maximum of 4.9~dB in the L-band, and by a mean of 1.3~dB and a maximum of 2.6~dB in the C-band, resulting in a throughput gain of 4.4~Tb/s. An example of the GLA transfer function applied to the channel centred at 1580.1~nm is shown in Fig.~\ref{fig:snr_tf}(b), where 400~ppm CO\textsubscript{2} reduces the peak attenuation by 32.3~dB.

We next investigated the impact of GLA suppression by gas purging in long-haul C+L-band HCF transmission. The C+L-band SMF results for different span lengths, each with the optimised flat launch power per band, are used as the benchmark. The corresponding throughput and launch power are listed in Table~\ref{tab:smf}.

\begin{table}[h!]
    \centering
    \small
    \caption{1000~km SMF results in C/L-band} \label{tab:smf}
    \begin{tabular}{|c|c|c|}
        \hline
        Span length & Throughput & Optimised power \\
        (km) & (Tb/s) & (dBm) \\
        \hline
        25 & 49.7 / 56.7 & 15 / 16 \\
        50 & 49.0 / 55.8 & 16.5 / 17.5 \\
        100 & 42.2 / 48.1 & 20 / 21 \\
        125 & 37.4 / 42.3 & 21.5 / 22.5 \\
        200 & 19.8 / 21.7 & 26.5 / 27.5 \\
        \hline
    \end{tabular}
\end{table}%

The next step was to compare the 1000~km transmission throughput of SMF and HCF, as shown in Fig.~\ref{fig:throughput_gla_span}. In this analysis, 96 scenarios were considered, covering 3 to 10 spans, four launch powers per band (27, 32, 36, and 40~dBm), and three residual gas conditions: 0, 400, and 1000~ppm CO\textsubscript{2}. In the C-band, the impact of GLA is relatively limited. At a 100~km span length, HCF with 27~dBm launch power achieves a throughput gain of 3.4~Tb/s over SMF, owing to its lower loss and nonlinearity. In this case, the number of repeaters along the link can be reduced by a factor of 2.7. Under ideal GLA suppression, the throughput gain in the C-band increased to 4.6~Tb/s at a 100~km span length.  Increasing the launch power further to 40~dBm, mainly benefits longer-span transmission, since shorter-span cases are limited by either IMI or transceiver noise. In this work, the IMI-induced SNR of 1000~km was fixed at 22~dB.

In the L-band, by contrast, GLA is the dominant performance-limiting factor. Without GLA suppression, SMF with an 8$\times$125~km configuration outperforms HCF with 100~km spans even at 40~dBm launch power, achieving 40.2~Tb/s. With 400~ppm CO\textsubscript{2}, L-band HCF provides an average throughput gain of 5.3~Tb/s, increasing the throughput to 45.5~Tb/s. Under ideal GLA suppression, the average throughput gain rises by 17.9~Tb/s across the different span lengths and launch powers, reaching a maximum of 59.2~Tb/s, which is approximately 1.5$\times$ higher. GLA suppression in the L-band also enables longer span lengths and therefore reduces the number of repeaters along the link. For 0~ppm CO\textsubscript{2}, up to 3.3$\times$ fewer repeaters than SMF link are required at a 27~dBm launch power. Even longer spans become feasible at higher launch powers, until performance is ultimately limited by IMI.


\section{Conclusions}
We investigated the impact of GLA on long-haul C+L-band HCF transmission using a numerical model with closed-form SMF results used as a benchmark. The results show that GLA has only a limited impact in the C-band but strongly limits the achievable throughput in L-band. With ideal suppression, the average L-band throughput gain in a 1000~km transmission link is 17.9~Tb/s, giving a 59.1~Tb/s throughput at 40~dBm launch power, an increase of approximately 1.5$\times$, while enabling up to 3.3$\times$ fewer repeaters at 27~dBm launch power. These results show that GLA suppression is essential for unlocking the full potential of future HCF-based C+L-band systems, although IMI remains the ultimate limitation once GLA is mitigated. In practical deployment, an additional challenge is that residual ambient gases may gradually re-enter the fibre after processing, potentially reintroducing absorption features and reducing the long-term stability of the achieved performance gains.

\clearpage
\section{Acknowledgements}
This work was supported by EPSRC Grant TRANSNET (EP/R035342/1), and EWOC (EP/W015714/1). Eric Sillekens is supported by Department for Science, 
Innovation and Technology and the Royal Academy of Engineering under the Research Fellowship scheme. Polina Bayvel is supported by a Royal Society Research Professorship.


\printbibliography

\vspace{-4mm}

\end{document}